\documentclass[pre,showpacs]{revtex4-1}
\usepackage{amsmath}
\usepackage{amsfonts}
\usepackage{amssymb}
\usepackage{graphicx}
\begin{document}
\title{Evolutionarily Stable Density-Dependent Dispersal}
\author{Shlomit Weisman}
\email{moshkovi@gmail.com}
\affiliation{Department of Physics, Bar Ilan University, Ramat-Gan 52900 Israel}
\author{Nadav M. Shnerb}
\email{shnerbn@mail.biu.ac.il}
\affiliation{Department of Physics, Bar Ilan University, Ramat-Gan 52900 Israel}
\author{David A. Kessler}
\email{kessler@dave.ph.biu.ac.il}
\affiliation{Department of Physics, Bar Ilan University, Ramat-Gan 52900 Israel}
\begin{abstract}
An ab-initio numerical study of the density-dependent, evolutionary
stable dispersal strategy is presented. The simulations are based on
a simple discretei generation island model with four processes:
reproduction, dispersal, competition and local catastrophe. We do
not impose any {\em a priori} constraints on the dispersal schedule,
allowing the entire schedule to evolve. We find that the system
converges at long times to a unique nontrivial dispersal schedule
such that the dispersal probability is a monotonically increasing
function of the density. We have explored the dependence of the
selected dispersal strategy on the various system parameters: mean
number of offspring, site carrying capacity, dispersal cost and
system size. A few general scaling laws are seen to emerge from the
data.
\end{abstract}
\maketitle

\section{Introduction}
Dispersal is one of the significant tools of survival. Like all biological characteristics, the dispersal rate is
determined by many factors. Starting from the groundbreaking work of Hamilton and May~\citep{HM,Hamilton2}, who introduced a simple
model of dispersal between islands, for which the evolutionary stable dispersal rate could be solved for as a function of the dispersal cost, many follow-on studies have been performed to understand the effect of area resources, survival rate of dispersal, local extinction probability and patch capacity on the dispersal rate.

As first discussed by \citet{olivieri}, one factor that can influence dispersal rates is local population density.  There is indeed evidence that dispersal rate increases with increasing population density, at least in invertebrates~\citep{invert1,invert2}.  Since then, there have been various attempts to determine the
evolutionary stable strategy (ESS). That this is a nontrivial exercise can be seen by the extremely complicated solution of \citet{Hamilton2} (CHM)
for the ESS in the much simpler
Hamilton-May model, where the migration rate is assumed to be independent of density.  Only in the limit of large carrying capacity does the CHM solution
reduce to a simple form, since the role of fluctuations is eliminated and each individual simply needs to maximize his expected number of offspring. Most
treatments to date have either assumed an ad-hoc parametric form for the dependence of dispersal rate on density, and calculated the evolutionary stable value of this parameter, or ignored the role of fluctuations to derive an analytical result.  \citet{art3} used an individual based simulation with an assumed linear
relationship between population density and dispersal probability.  \citet{art6}, \citet{art1} and \citet{art5} studied more complicated functional relationships between density and dispersal rate,
 with a threshold density below which there was no dispersal. Metz $\&$ Gyllenberg~\citep{metz,metz1} claimed to prove that, at least in the limit of large
 carrying capacity, that the ESS has the form of a
step-function, with a population density threshold below which individuals do not migrate and above which they all
migrate.  We will return to this in the discussion.

In this paper, we study how the population density affects the dispersal rate using a variant of the original
Hamilton-May model~\citep{Hamilton2}, a simple discrete generation
island model with four processes: reproduction, dispersal, competition and local catastrophe. In contrary to the
Hamilton-May type model, where each individuals has some given dispersal probability, $\nu$, in our system each
individual is characterized by a dispersal probability for every possible local patch density,  $\nu(d)$.  The system
undergoes an evolutionary process wherein each individual descendant inherits its parent's dispersal schedule,
subject to small mutational modification.  Thus, as opposed to the previous works on this problem, we do not
constrain the dispersal schedule, $\nu(d)$, to have any a priori given form.  We find that the system converges at
long times to a unique nontrivial dispersal schedule $\nu^*(d)$. In general, this $\nu^*(d)$ is a monotonically
increasing function of $d$, but in general does not seem to follow any simple functional form.  We have explored the
 dependence of the evolutionarily selected $\nu^*(d)$ on the various system parameters, namely the mean number of
offspring, the site carrying capacity, the dispersal cost and the system size.

\section{The model}
\begin{figure}[h]
\includegraphics[width=0.3\textwidth]{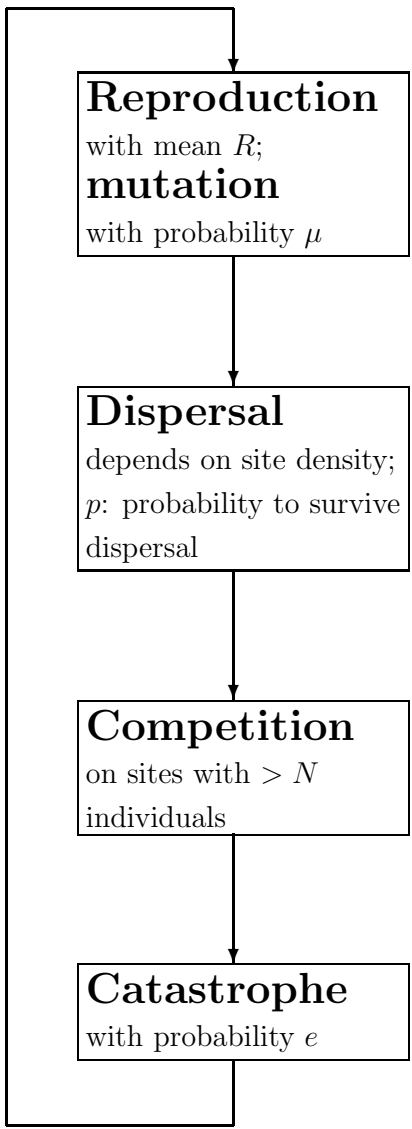}
\caption{Cycle of events in each generation.}
\label{fig:cycle}
\end{figure}

The model (see Fig.\ref{fig:cycle}) is a discrete generation island model with four successive processes during each generation: reproduction,
dispersal, competition and local catastrophe. The dispersal genetic information is an array of dispersal rates
for each possible  number of individuals in the local site before dispersal.

The system is made up of a very large number of sites each capable of supporting at most $N$ reproductive adults. Each adult
independently produces a number of offspring according to a Poisson distribution with mean $R$.   After the reproduction stage, all the mature individuals
die. During dispersal, each offspring independently disperses with probability $\nu(d)$, according to the occupancy $d$
of the source site. Migrants from all sites enter a pool and have a probability $p$ to survive the
dispersal. The survivors are then distributed in a uniform random manner among the sites.

Each site has a probability $e$ to undergo a local catastrophe that kills all the individuals on the
site.  Barring such a catastrophe,
competition then occurs between all the current occupants of every given overloaded site with occupancy greater than the carrying capacity $N$, both the non-dispersing offspring and the immigrants. $N$ survivors are chosen at random from the competitors.
Sites with less than $N$ individuals skip the
competition stage. This is the essential difference between the present model and that of Hamilton and May, where underpopulated sites were restored to
full capacity during the ``competition'' phase.  This, in addition to the noisy birth process, is what produces a wide range of densities in the model, since after a local catastrophe, it takes a number of cycles for the site to become fully populated again.

Each offspring has
 probability $\mu$ of having a mutation in one of his dispersal values. The new dispersal probability is determined according to a
binomial distribution with mean equal to $k$ times  the previous dispersal probability;  the new dispersal probability is the  resulting random number  divided by $k$. This has the important property that the mutation process is unbiased, with the mean new dispersal probability equal to the previous value.  Also, the new dispersal probability is guaranteed to fall between 0 and 1, as it should.  The size of the typical change in dispersal probability is controlled by the parameter $k$.  In practice, we have taken $k=1000$ in our simulations.

\section{Results}
\begin{figure}[h]
\includegraphics[width=0.8\textwidth]{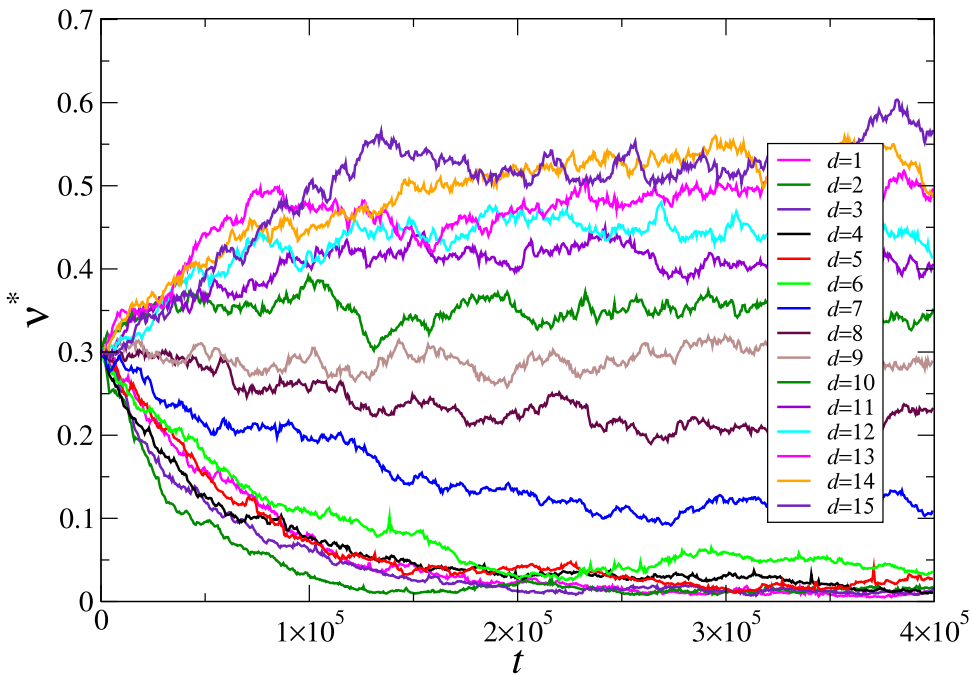}
\caption{Population average dispersal rate as a function of time for different local densities, $d$. $N=5$, $p=0.2$, $e=0.1$, $\mu=0.01$, $R=2$.}
\label{fig:time}
\end{figure}

For all parameter combinations tested in the simulations, the initial dispersal rate $\nu(d)$ was taken to be a
constant. The population at long time converged to a  evolutionarily selected average dispersal rate schedule, $\nu^*(d)$. The
reported results for $\nu^*(d)$ are the average of ten systems which ran over
700000 generations after the system reached stable values, as depicted in Fig. \ref{fig:time} for the parameter set $p=0.2$, $N=5$, $e=0.1$, $R=2$, $\mu=0.01$. The system contain 1000 sites. The
dispersal schedule $\nu^*(d)$ was calculated for those values of the site population  that are realistic for the site
 to attain. This function $\nu^*(d)$ is to be compared with the value $\nu_0^* \approx 0.253$ (for the above parameter set) when $\nu_0$ is
 constrained to be independent of $d$

\begin{figure}[ht]
\includegraphics[width=0.45\textwidth]{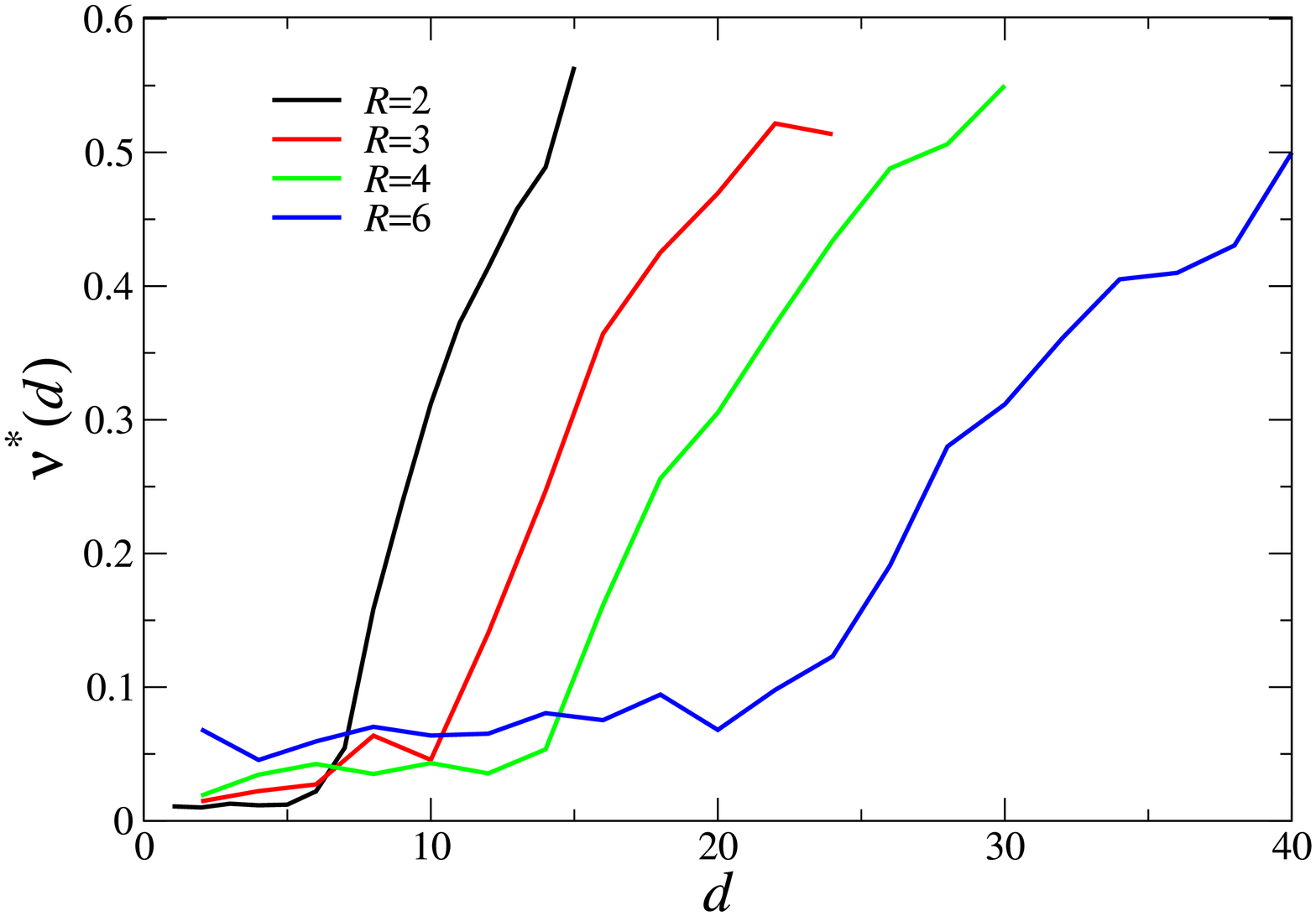}
\includegraphics[width=0.45\textwidth]{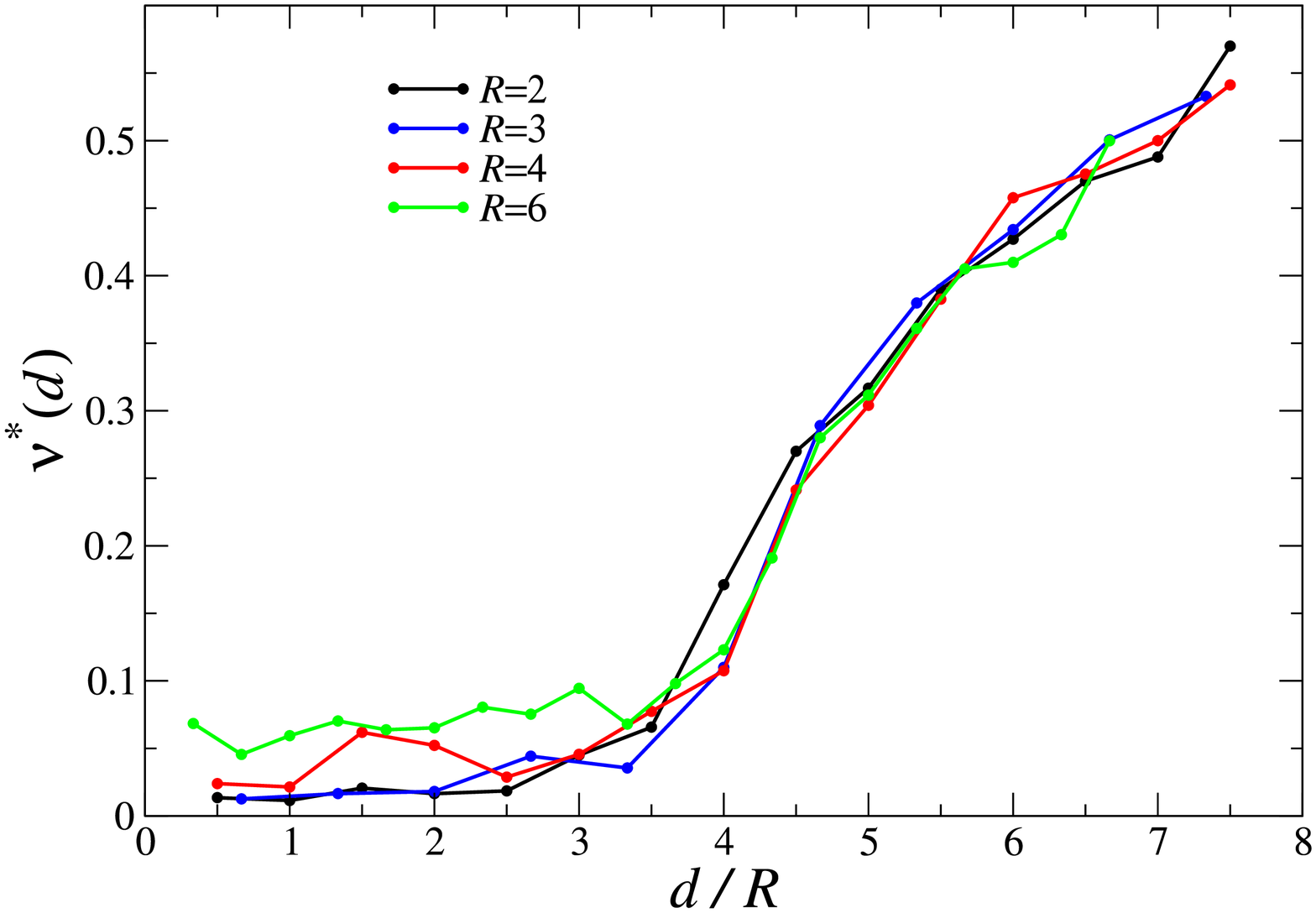}
\caption
{Dispersal rate for different mean reproduction rates, $R$. $p$=0.2, $e$=0.1, $\mu$=0.01, $N=5$.   Left:  Dispersal
rate as a function of density. Right: Dispersal rate as a function of scaled density, $d/R$.}
\label{fig:Time_R}
\end{figure}

The resulting stable dispersal schedule $\nu^*(d)$ is shown in the
left-hand panel of Fig. \ref{fig:Time_R}, for various values of the
reproduction rate (i.e., average number of offspring per
individual), $R$. One striking feature is that as long as the
population density remains below some threshold density, very little
emigration occurs. This threshold density is determined not solely
by the carrying capacity $N$, as one might assume, but rather is
given by the average site occupancy before dispersal. This can be
seen very clearly by plotting $\nu^*$ versus the scaled density
$d/R$, as shown in the right-hand panel. In fact, above the
threshold, all the curves coincide so that the selected dispersal
rate is essentially a function only of this scaled density. This
scaling of the threshold can be understood be noting that below the
threshold. i.e., for $d<R$, there is nothing gained in leaving the
site. On the contrary, there is a lower chance of a migrating
individual to stay alive because most likely he will migrate to a
more highly occupied site. Above the threshold, the site is
overcrowded relative to the average density, so migration increases
the chance of staying alive. The dispersal probability below the
threshold is seen to increase slightly as the reproductive mean
increases.

Fig. \ref{fig:Time_R} also illustrates an intriguing result, namely
that the higher the mean reproduction rate, the smaller the
dispersal probability. This is somewhat counter-intuitive, since one
would naively assume that higher fecundity would result in a surplus
population at home which is better used in attempted colonization of
empty sites. The result is very interesting because its suggests an
evolutionary  mechanism that leads to a colonization/competition
tradeoff \citep{compcol1,compcol2}.

\begin{figure}[h]
\includegraphics[width=0.45\textwidth]{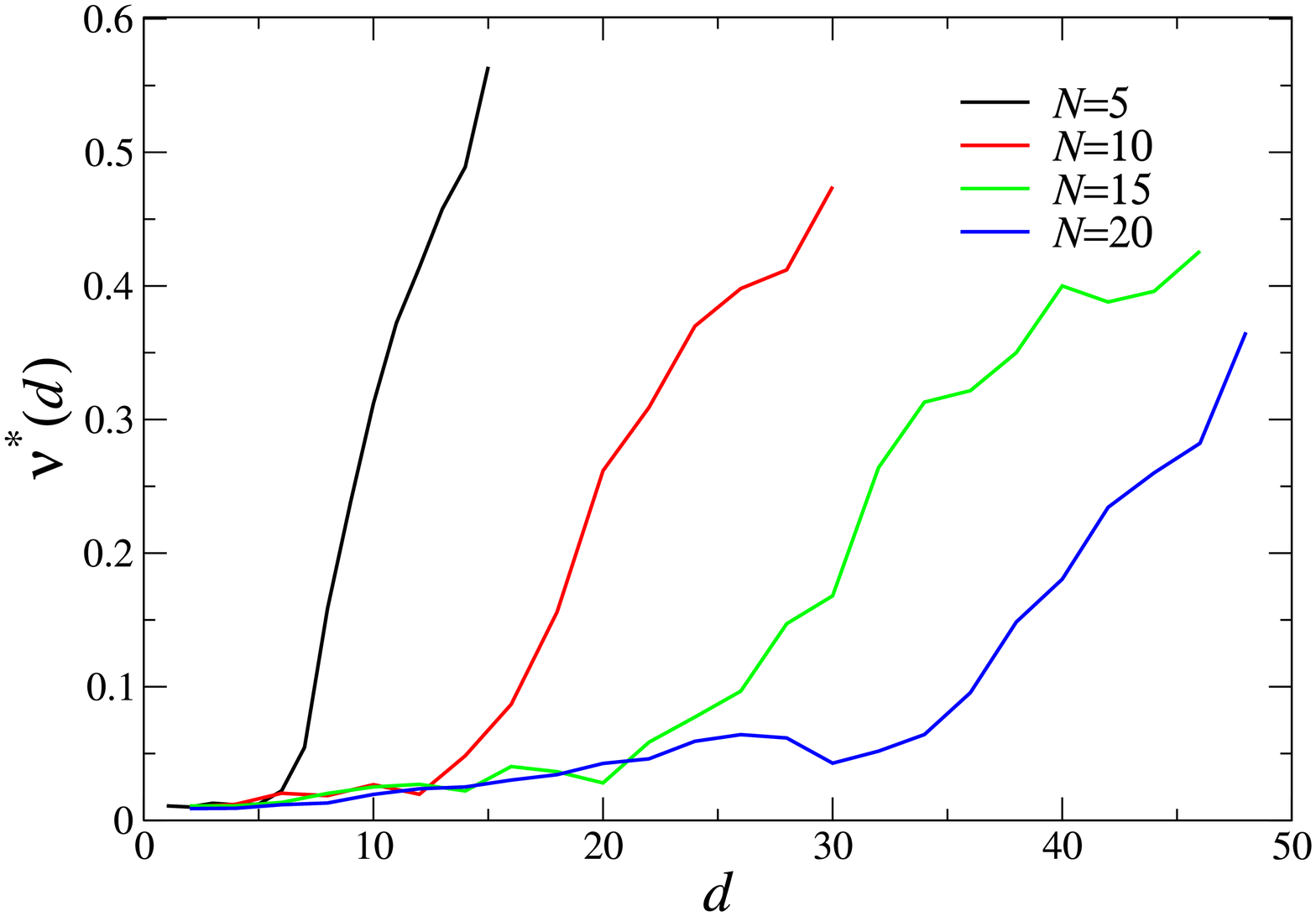}
\includegraphics[width=0.45\textwidth]{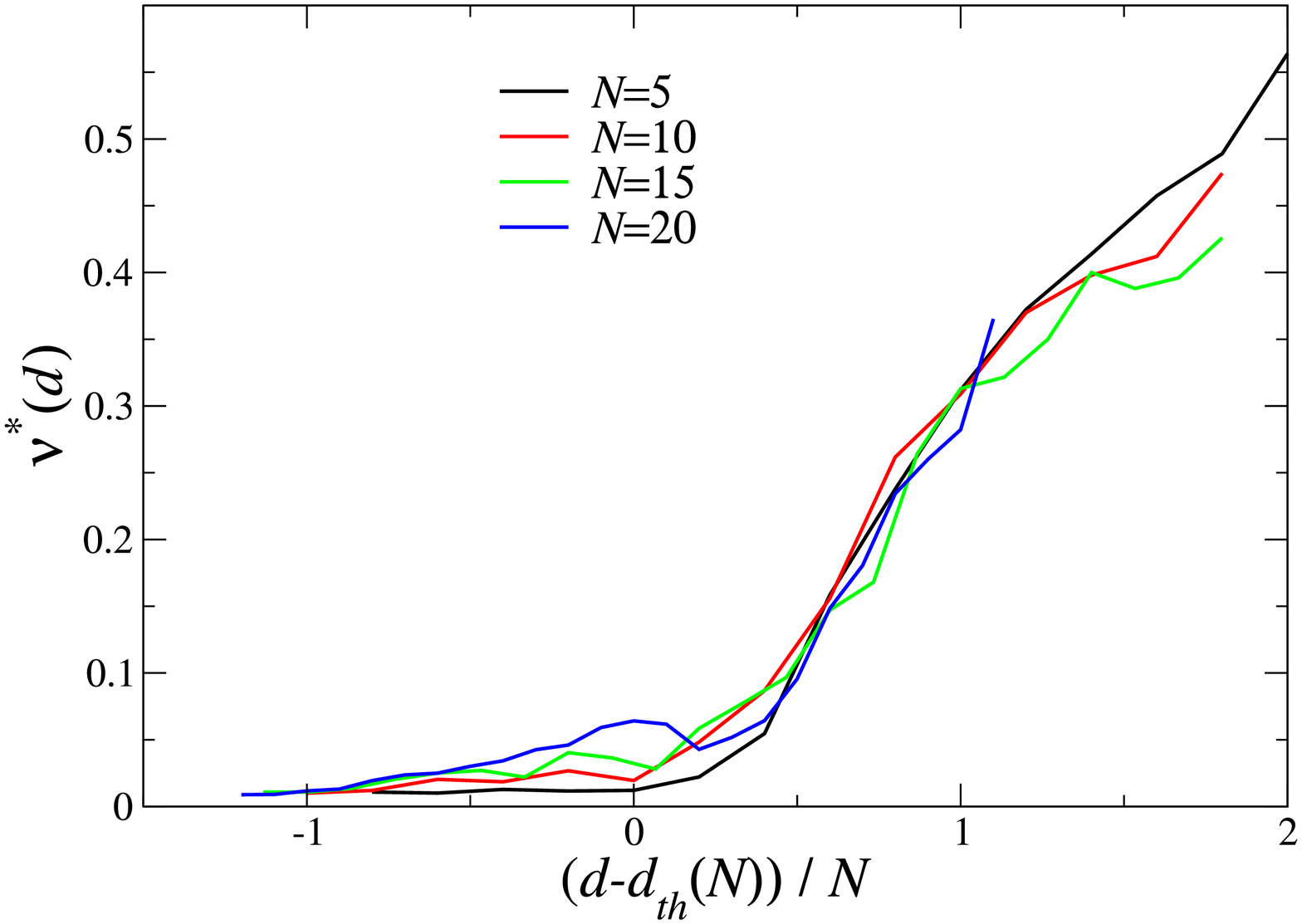}
\caption{Dispersal rate for different values of the carrying capacity, $N$.  $N$. $p=0.2$, $e=0.1$, $\mu$=0.01,
$R=2$. Left: Dispersal rate as a function of density, $d$. Right:  Dispersal rate as a function of scaled density, $(d-d_{\mathit{th}}(N))/N$, showing collapse.  Here $d_{\mathit{th}}(N) = 1.4N - 2$.}
\label{fig:time_n_all}
\end{figure}

The site carrying capacity, $N$, also effects the dispersal as can
be seen in Fig. \ref{fig:time_n_all}. As the number of surviving
individuals after competition grows, the threshold density increases
apparently linearly, as $d_{\mathit{th}}(N)\approx 1.4N-2$, for
$R=2$, $p=0.2$, and $e=0.1$.  Above threshold, the slope of the
dispersal curve decreases with increasing $N$. Here again, as with
$R$, the data can be effectively collapsed, by plotting $\nu^*$ as a
function of $(d-d_{\mathit{th}}(N))/N$.   The general trend that
$\nu^*$ decreases with $N$ is  consistent with its behavior  in the
standard density-independent Hamilton-May model.  Here, the
subthreshold dispersion probability  rises slightly with $N$.

The last parameter that was investigated is $p$, the probability to survive dispersal.
Dispersal rates increases and threshold density decreases with increasing $p$.  It should be noted that there is no sign of the nonmonotonic behavior of $\nu^*$
with $p$ seen in the density-independent model.
\begin{figure}[h]
\includegraphics[width=0.8\textwidth]{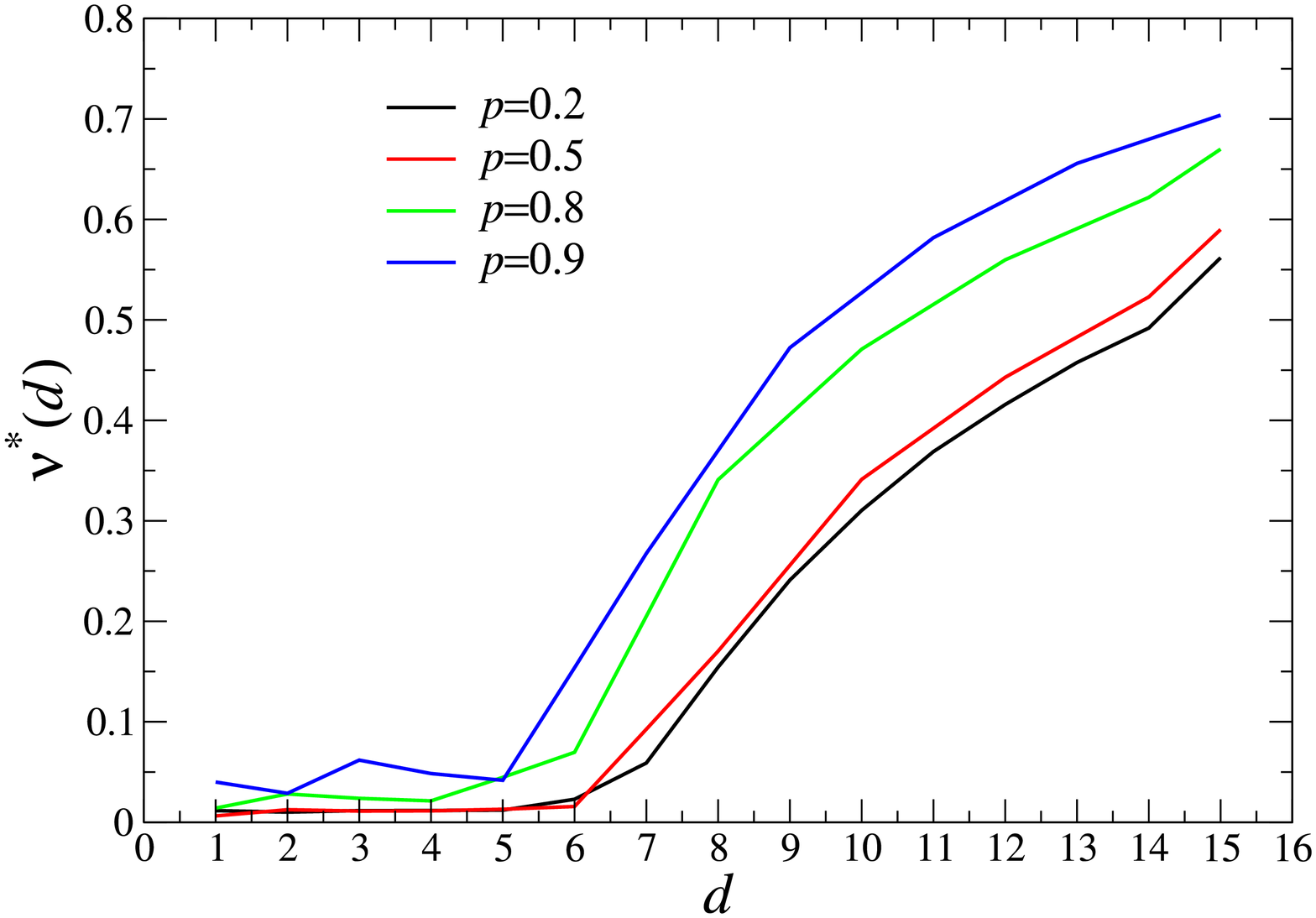}
\caption{Dispersal rate as a function of density for different probabilities to survive dispersal, $p$. $N=5$, $e=0.1$,
$\mu$=0.01, $R$=2}
\label{fig:time_p}
\end{figure}

\begin{figure}[h]
\includegraphics[width=0.8\textwidth]{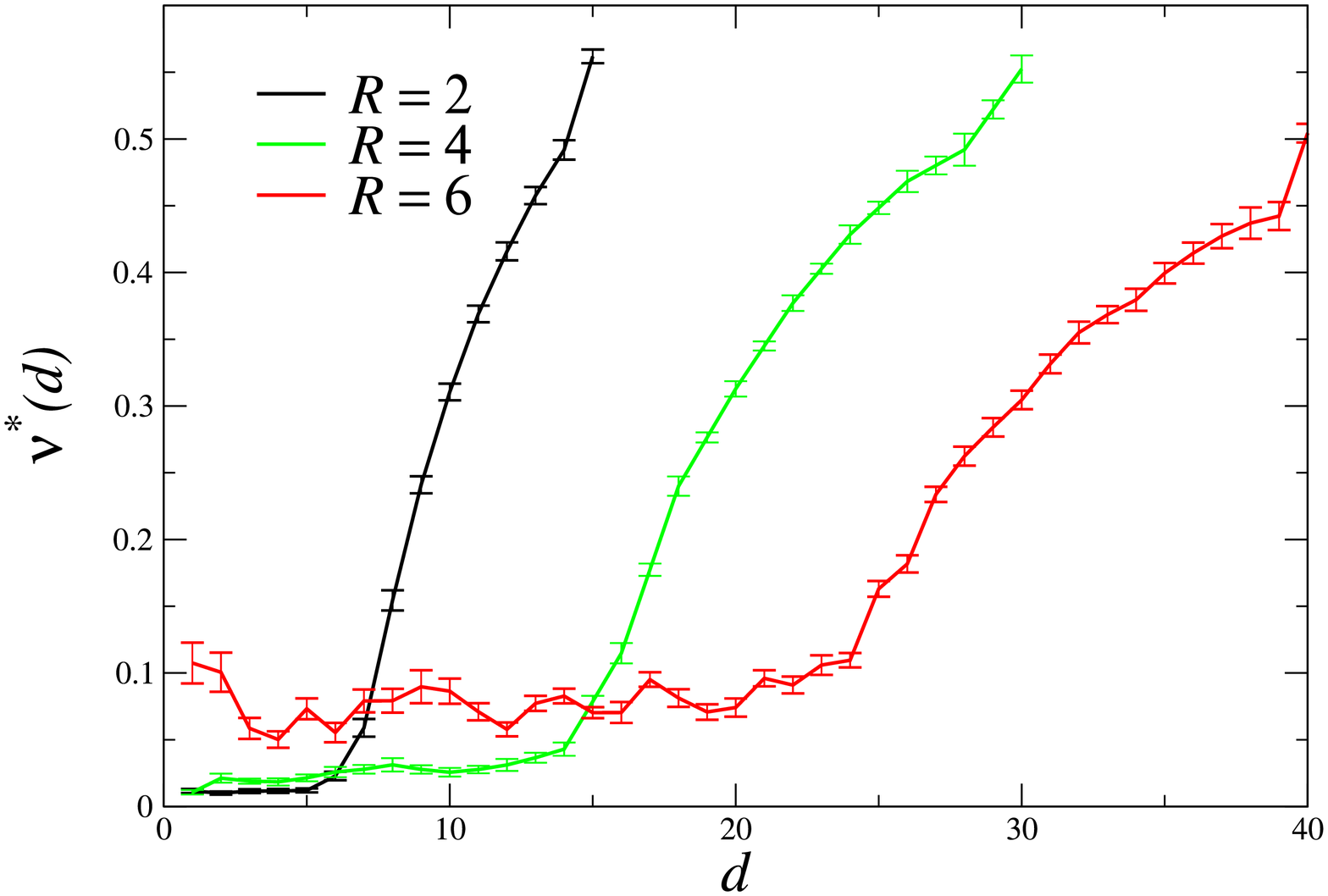}
\caption{High precision calculation of dispersal rate as a function of density for different mean reproduction numbers, $R$, showing the standard error. $N=5$, $e=0.1$,
$\mu=0.01$, $p=0.2$}
\label{fig:std}
\end{figure}

In Fig. \ref{fig:std}, we present much longer runs in order to clean up the data and in addition measure the standard error of the
results for $R=2,4,6$. For $R=6$ the results are averaged over sixty systems and are noisier mainly in the area
below the threshold.

\section{Discussion}
In this work we have evolved via simulation the evolutionary stable strategy, where the dispersal rate is
sensitive to population density. Previous studies in this field restricted  the dispersal function to a predetermined functional
\citep{art3,art1} with a single evolving parameter. \citet{art5} used a more complicated function with three parameters that can evolve. The current study allows for a general dispersal schedule, $\nu^*(d)$, without a priori constraints.

We studied the evolution of density-dependent dispersal in a simple discrete generation island model with four
processes (reproduction, dispersal, competition and local catastrophe). The dispersal genetic information is an
array of dispersal rates for each possible number of individuals in the local site before dispersal. The dispersal
rates change by mutation, and  converge to a nontrivial function.

One noteworthy characteristic is the absence of any sign of an all-or-nothing strategy emerging, in contrast to the predictions of Metz and Gyllenberg~\citep{metz,metz1}.  It should be pointed out that the field data on butterflies~\citep{butterfly} also shows a smooth increase in dispersal with local density.
What then distinguishes the model of Metz and Gyllenberg? One possibility is that their models are formulated in continuous time, whereas time in the current model is discretized in units of reproduction cycles.  Our model can be forced to approach the continuous time limit by letting $R$, the reproduction number approach unity.  Indeed, decreasing $R$ increases the slope of the $\nu^*(d)$ curve, as seen in Fig. 3.  Thus, it is reasonable to conjecture that in the limit of $R \to 1$, the slope does approach infinity, in accord with the Metz and Gyllenberg results. 

 In any case, the field data on butterflies shows no sign of a region of no dispersal for very low densities, which is a universal outcome of our simulations.  It is possible that the densities encountered in the study where all above the threshold.  This point merits more study.

\bibliography{bibfile2}

\end{document}